\documentclass[11pt,twoside]{article}

\usepackage{graphicx}
\usepackage{asp2006}
\usepackage{epsf}
\usepackage{psfig}

\begin{document}
\title{ATCA Detection of SiO Masers in the Inner Parsecs of the Galactic Center}   

\vspace{-5mm}

\author{Juan Li\altaffilmark{1,3}, Tao An\altaffilmark{1}, Zhi-Qiang Shen\altaffilmark{1,2}, Atsushi Miyazaki\altaffilmark{4}}

\affil{1 Shanghai Astronomical Observatory, Chinese Academy of
Sciences, Shanghai 200030, China; lijuan@shao.ac.cn} \affil{2 Joint
Institute for Galaxy and Cosmology (JOINGC) of SHAO and USTC,
Shanghai 200030, China;} \affil{3 Graduate School of the Chinese
Academy of Sciences, Beijing 100039, China}
 \affil{4 National Astronomical
Observatory of Japan, 2-21-1 Osawa, Mitaka, Tokyo 181-8588, Japan}

\vspace{-5mm}

\begin{abstract} We present sensitive Australia Telescope Compact Array (ATCA)
observations of SiO masers in the inner parsecs of the Galactic
center (GC). We detected five SiO J=2-1, v=1 (86 GHz) masers in the
innner 25$^{\prime\prime}$ (1 pc) of the GC. All the detected 86 GHz
SiO masers are previously known SiO J=1-0, v=1 (43 GHz) masers and
associated with late-type stars. Eighteen 43 GHz SiO masers were
detected within 50$^{\prime\prime}$ (2 pc) of Sagittarius A*. Among
them, seven are detected for the first time, which brings the total
number of 43 GHz SiO masers to 22 in this region. 
\end{abstract}


\vspace{-5mm}

SiO maser emission arising from the circumstellar envelope is
believed to be a reliable tracer of the gravitational field because
they can be treated as point-like particles and are not subject to
forces from magnetic fields, winds, or collisions. At 43 GHz,
fifteen SiO masers have been detected with VLA (Menten et al. 1997; Reid et
al. 2003, 2007). However, these observations were limited by the receiver band coverage. Limited by
observing facilities, the number and distribution of 86 GHz SiO
maser sources in the central few parsecs of the GC is poorly known
(Sjouwerman et al., in prep.).

At 86 GHz, we observed Sgr A* and its vicinity with Australia
Telescope Compact Array (ATCA) of the Australia Telescope National
Facility (ATNF) in October, 2007.
Two data sets were acquired from the ATCA archive, which
were performed in 2006 May and 2007 May.
Our 43 GHz ATCA observations were conducted on June 3 and October 3,
2008.
All the observations pointed at the position of Sgr A*,
allowing detection of masers within the primary beam of an ATCA
antenna ($\sim 36^{\prime\prime}$ FWHM at 86 GHz, $\sim
72^{\prime\prime}$ FWHM at 43 GHz). The data were reduced in the
standard manner using MIRIAD.

At 86 GHz, five maser sources associated with IRS 7, IRS 12N, IRS
28, IRS 10EE and IRS 15NE were detected. IRS 7 is likely to be the
strongest 86 GHz SiO maser source in the inner 1 pc of the GC
region, with a peak flux density of about 641 mJy B$^{-1}$ in May 2006. Eighteen 43
GHz SiO maser sources were detected. Eleven of them are identified
within the errors with previous detections. Seven new SiO maser
stars were discovered. One of these new detections, offset from Sgr
A* by (+0.$^{\prime\prime}$9, -8.$^{\prime\prime}$0), is coincident
within the errors with IRS 14NE, a known AGB star (Blum et al.
1996). Following the naming of SiO maser sources adopted by Reid et
al. (2007), other six new detections were named as SiO-18, SiO-19,
SiO-20, SiO-21, SiO-22, SiO-23, respectively. This has brought the
total number of 43 GHz SiO masers to 22.
Fig. 1 shows the 7 mm ATCA
continuum map of the inner 50$^{\prime\prime}$ of the GC region.
Positions of maser stars are marked with their LSR velocities and IRAS names indicated in the map.

\begin{figure}[!ht]
\begin{center}
\includegraphics[scale = 0.5]{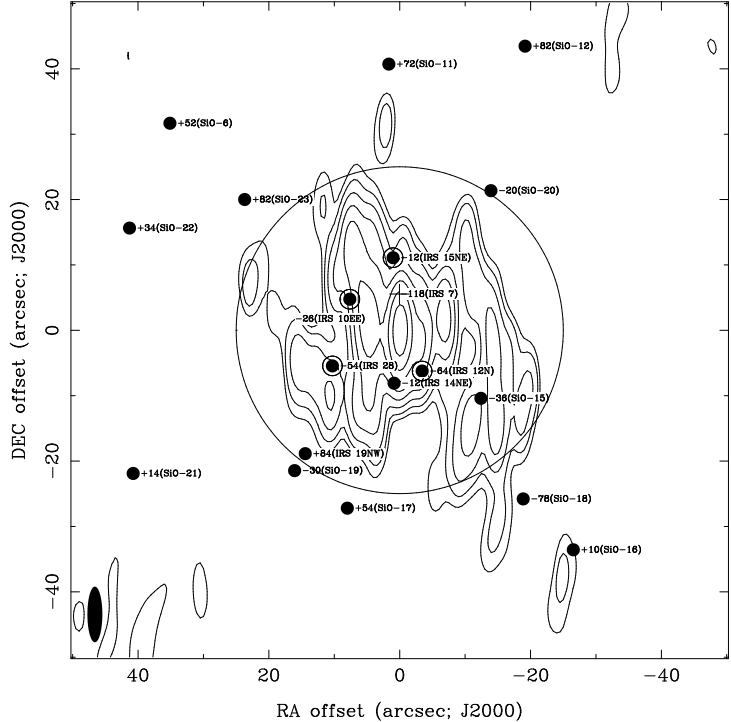}
\caption{The 7 mm ATCA continuum map of the GC region. The size of
the region corresponds to an area of projected size 4 pc $\times$ 4
pc centered on the position of Sgr A*. Late-type giant and
supergiant stars with 43 GHz SiO maser only (solid circle), 86 GHz SiO
maser only (cross) and both 43 and 86 GHz SiO maser emission (solid and open circle) are marked. A circle
represents the central 1 pc (25$^{\prime\prime}$) region.}\label{fig2}
\end{center}
\end{figure}

\vspace{-7mm}

\acknowledgements 
The Australia Telescope Compact Array is part of the Australia
Telescope which is founded by the Commonwealth of Australia for
operation as a National Facility managed by the CSIRO.
This work was supported in part by the National Natural
Science Foundation of China (grants 10625314, 10633010 and
10821302) and the Knowledge Innovation Program of the Chinese
Academy of Sciences (Grant No. KJCX2-YW-T03), and sponsored by the
National Key Basic Research Development Program of China (No.
2007CB815405) and the Ministry of Science and Technology of China (Grant No. 2009CB824900/2009CB24903). 

\vspace{-4mm}


\end{document}